\documentclass[10pt,twocolumn,letterpaper]{article}

\usepackage[pagenumbers]{cvpr}  

\usepackage[accsupp]{axessibility}

\usepackage{marvosym}
\usepackage{makecell}
\usepackage{multirow}
\usepackage{algorithm}
\usepackage{algorithmic}

\definecolor{cvprblue}{rgb}{0.21,0.49,0.74}
\definecolor{revisionblue}{rgb}{0.0,0.0,1.0}

\usepackage[pagebackref,breaklinks,colorlinks,allcolors=cvprblue]{hyperref}

\def\paperID{5798}
\def\confName{CVPR}
\def\confYear{2026}

\title{Modeling Spatiotemporal Neural Frames for High Resolution Brain Dynamics}

\author{Wanying Qu$^{1}$, Jianxiong Gao$^{1}$, Wei Wang$^{2,3}$, Yanwei Fu$^{1,3}$$^{\text{\Letter}}$\\
$^{1}$Fudan University, $^{2}$Southern University of Science and Technology, $^{3}$Shanghai Innovation Institute\\
{\tt\small \{wyqu24,jxgao22\}@m.fudan.edu.cn,\ \ 12445027@mail.sustech.edu.cn, \ \ yanweifu@fudan.edu.cn}}

\begin{document}
\twocolumn[{
\maketitle

\begin{center}
\includegraphics[width=0.8\linewidth]{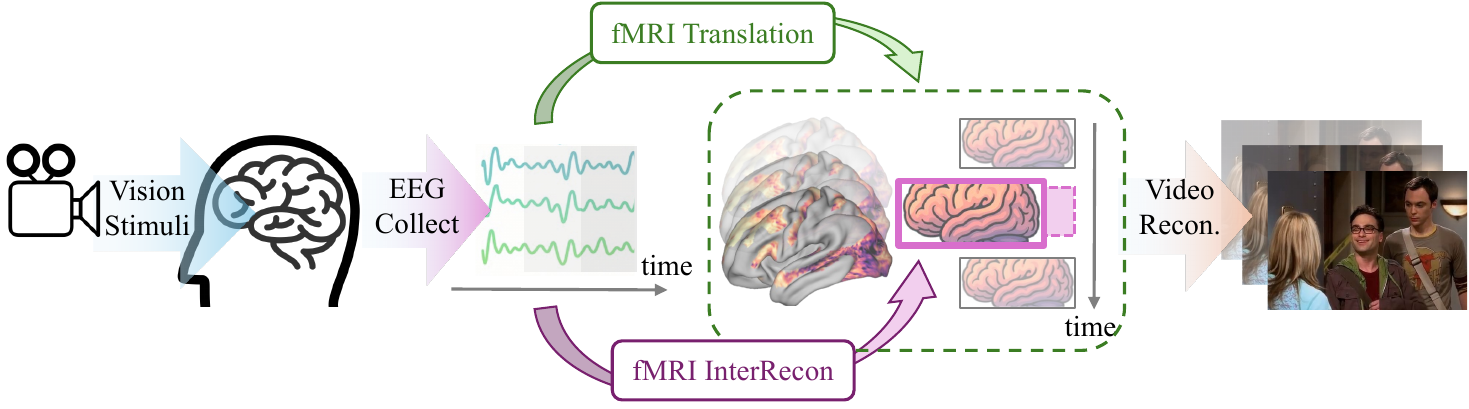}
\end{center}
\vspace{-0.7cm}
\captionsetup{type=figure}
\captionof{figure}{%
This paper reconstructs dynamic fMRI frames with high spatial detail and temporal coherence from EEG. We further introduce intermediate frame reconstruction (InterRecon), and use visual decoding as a downstream task to evaluate high-level semantic preservation.
}\label{fig:architecture}
\vspace{0.3cm}
}]
{\renewcommand{\thefootnote}{}\footnotetext{$^{\text{\Letter}}$Corresponding author. }}
\begin{abstract}
Capturing dynamic spatiotemporal neural activity is essential for understanding large-scale brain mechanisms. Functional magnetic resonance imaging (fMRI) provides high-resolution cortical representations that form a strong basis for characterizing fine-grained brain activity patterns. The high acquisition cost of fMRI limits large-scale applications, therefore making high-quality fMRI reconstruction a crucial task. Electroencephalography (EEG) offers millisecond-level temporal cues that complement fMRI. Leveraging this complementarity, we present an EEG-conditioned framework for reconstructing dynamic fMRI as continuous neural sequences with high spatial fidelity and strong temporal coherence at the cortical-vertex level. To address sampling irregularities common in real fMRI acquisitions, we incorporate a null-space intermediate-frame reconstruction, enabling measurement-consistent completion of arbitrary intermediate frames and improving sequence continuity and practical applicability. Experiments on the CineBrain dataset demonstrate superior vertex-wise reconstruction quality and robust temporal consistency across whole-brain and functionally specific regions. The reconstructed fMRI also preserves essential functional information, supporting downstream visual decoding tasks. This work provides a new pathway for estimating high-resolution fMRI dynamics from EEG and advances multimodal neuroimaging toward more dynamic brain activity modeling.
\end{abstract}
\section{Introduction}
\label{sec:intro}

Understanding brain dynamics requires capturing both high spatial detail and temporal continuity~\cite{poldrack2012future}. Functional magnetic resonance imaging (fMRI) measures neural activity through blood-oxygen-level–dependent (BOLD) signals and provides fine-grained spatial resolution for studying cognition and brain function~\cite{logothetis2001neurophysiological,glover2011overview}. Each fMRI frame reflects an ongoing state of brain activity, and together these frames form a continuous temporal sequence. The high acquisition cost of fMRI limits large-scale applications, therefore making high-quality fMRI reconstruction a crucial task. However, conventional analyses and many reconstruction methods treat frames independently~\cite{lindquist2008statistical,calhas2022eeg, catd}, overlooking the rich temporal dependencies that govern how neural patterns emerge, evolve, and interact. Modeling these dynamics is inherently challenging, as it requires reconstructing high-dimensional spatiotemporal sequences that remain both temporally coherent and spatially detailed.

This raises a central question: \textit{\textbf{how can we reconstruct fMRI frames that are both temporally coherent and spatially detailed, capturing the brain's continuous dynamics rather than isolated snapshots?}} To address this challenge, we turn to electroencephalography (EEG), which captures neural activity with millisecond temporal resolution~\cite{michel2012towards}, complementing the spatial precision of fMRI. By leveraging EEG as a conditioning signal, we can guide the reconstruction of dynamic fMRI frames that preserve temporal continuity across frames. Beyond reconstruction fidelity, a second critical question arises: \textit{\textbf{how to evaluate the reconstructed fMRI, and more importantly, whether it can support meaningful downstream applications.}}

Existing approaches only partially address these challenges. Methods focusing on temporal resolution, such as ROI-based models~\cite{li2024neurobolt} capture dynamic trajectories but operate at coarse spatial granularity, limiting interpretability for cortical mapping. Methods focusing on spatial resolution, including voxel-level renconstruction~\cite{liu2019convolutional,calhas2022eeg,roos2025brainwaves} and vertex-based fMRI translation models~\cite{catd} preserve spatial detail but lack temporal coherence, leading to framewise artifacts and biologically implausible inconsistencies.
Moreover, conventional evaluation paradigms rely on voxel or vertex level metrics such as MSE, SSIM~\cite{wang2004image}, and PSNR~\cite{hore2010image}, which cannot assess whether the reconstructed fMRI encodes meaningful neural representations usable for higher-level tasks.

To overcome these limitations, we propose a diffusion-transformer-based framework for EEG-conditioned fMRI reconstruction that addresses both temporal coherence and spatial resolution, as shown in Figure \ref{fig:architecture}. Our key insight is to model brain activity as evolving spatiotemporal neural frames rather than independent snapshots, jointly capturing vertex-level spatial detail and temporal continuity within a diffusion-based generative process. Critically, we introduce a null-space constrained sampling mechanism that enables intermediate frame reconstruction (InterRecon) without retraining. InterRecon serves two important purposes. First, it provides an intrinsic evaluation of temporal consistency, allowing us to assess whether the model can generate physiologically plausible transitions between known fMRI frames; Second, it addresses practical needs in real fMRI data processing, where missing or corrupted frames often occur during acquisition and interpolation is routinely required during preprocessing~\cite{allen2022massive}. Beyond reconstruction metrics, we validate functional plausibility through downstream visual decoding task, ensuring the reconstructed fMRI preserves task-relevant neural representations.

We demonstrate that this framework achieves both goals: reconstructing temporally coherent fMRI frames with high spatial fidelity while preserving functional information validated through downstream decoding tasks.
Our contributions are threefold:
\begin{itemize}
    \item We formulate EEG-to-fMRI reconstruction as a spatiotemporal pattern recognition and generation task, where the goal is to identify task-relevant neural dynamics from EEG and reconstruct corresponding fMRI frames. We introduce a diffusion transformer framework that reconstructs full fMRI sequences at vertex-level cortical resolution while explicitly modeling temporal dependencies and cortical spatial geometry to produce coherent and physiologically consistent representations. 
    \item We develop a null-space constrained sampling mechanism to reconstruct intermediate frames at arbitrary temporal positions without retraining, offering practical utility in real fMRI processing scenarios.
    \item We conduct a comprehensive evaluation of the reconstructed fMRI, assessing quantitative metrics such as vertex-level fidelity, temporal coherence, and structural correspondence, as well as functional validity through a downstream visual decoding task.
\end{itemize}
\section{Related Work}
\label{sec:formatting}

\subsection{EEG-to-fMRI Translation}

EEG-to-fMRI translation methods can be divided into ROI-level sequence modeling and voxel or surface-level reconstruction. ROI-based methods such as NeuroBOLT~\cite{li2024neurobolt} reconstruct fMRI sequences at the parcel level. These methods naturally capture temporal continuity across frames but lose spatial precision because parcellation removes fine cortical detail. Voxel-level or cortical-level approaches, such as CNN-TC~\cite{liu2019convolutional}, CNN-TAG~\cite{calhas2022eeg}, adversarial models including E2fGAN and E2fNet~\cite{roos2025brainwaves}, and CATD~\cite{catd} operate in volumetric or cortical space. They achieve higher spatial fidelity but typically reconstruct each fMRI frame independently. As a result, these methods lack explicit modeling of temporal coherence across the sequence.

Evaluation protocols follow the same pattern. ROI-level methods mainly report correlation-based metrics between predicted and ground-truth time series. Voxel-level methods commonly rely on image reconstruction metrics such as MSE, correlation, SSIM, and PSNR. These metrics reflect reconstruction fidelity but do not reveal whether the reconstructed fMRI contains functional information that can support downstream applications such as intermediate-frame completion or neural decoding.

Our work addresses these gaps. We introduce an EEG-conditioned diffusion framework that models spatial detail and temporal consistency jointly at vertex-level resolution. We further incorporate a null space-based strategy that enables measurement-consistent reconstruction of intermediate frames without retraining. More importantly, we evaluate whether the reconstructed fMRI preserves task-relevant neural representations that support downstream functional applications.

\subsection{Diffusion Models}
Denoising Diffusion Probabilistic Models (DDPMs)~\cite{ho2020denoising,sohl2015deep} have demonstrated strong capabilities in modeling complex data distributions, achieving remarkable success in image generation~\cite{podell2023sdxl}. Over time, diffusion models have evolved into a flexible probabilistic modeling framework whose iterative denoising process naturally captures multi-scale spatial structure and complex data manifolds while offering stable training dynamics. Subsequent works, such as Diffusion Transformers (DiT), have further scaled up diffusion architectures, enabling a new wave of powerful generative vision models~\cite{hong2022cogvideo,wan2025,longvie,longvie2}. This expansion has revealed diffusion’s strong capacity for high-dimensional sequence modeling, controllable generation, and modality-conditioned synthesis. Beyond visual synthesis, diffusion models have also been extended to cross-modal distribution modeling, including brain-to-vision generation from fMRI signals~\cite{mind-vis,mind-video,mind-3d, mind3d++}, reflecting their suitability for learning structured relationships across heterogeneous data domains.
\section{Method}

\begin{figure*}[t]
\centering
\includegraphics[width=\textwidth]{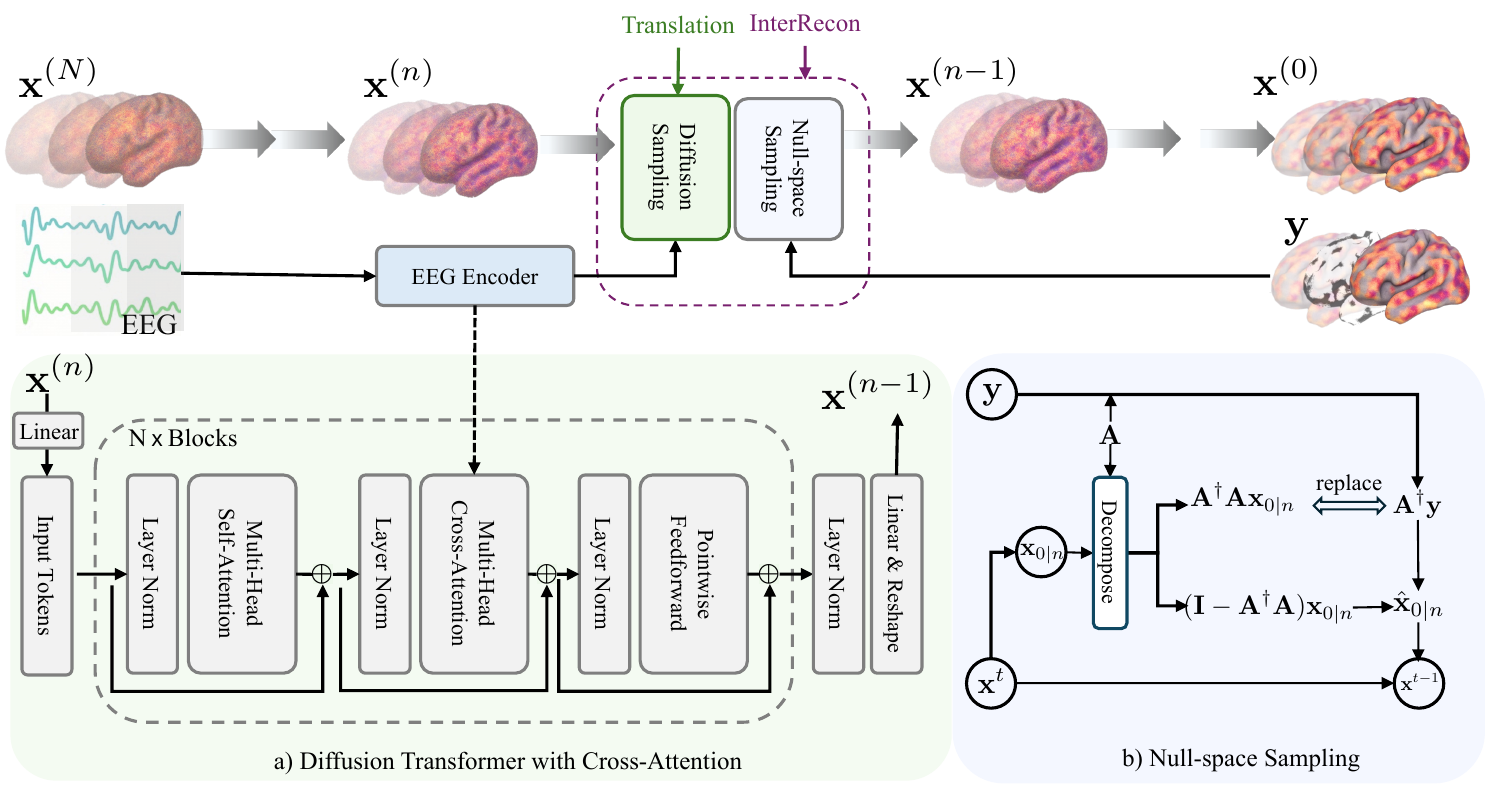}
\caption{Architecture and inference workflow of our framework.
Top: End-to-end inference pipeline, where EEG features condition a diffusion-transformer-based reconstruction process to produce spatiotemporal fMRI sequences.
Bottom: (a) Transformer-based denoising network that jointly models vertex-level tokens with sequence-level EEG cross-attention to reconstruct coherent fMRI trajectories. (b) Null-space sampling module (InterRecon) that enforces measurement consistency during reverse diffusion, enabling intermediate frame reconstruction without retraining.}
\label{fig:framework}
\end{figure*}

\subsection{Task Formulation}

In this study, we aim to construct a generative framework that learns to reconstruct fMRI frames from EEG measurements.

Let \(\mathbf{X}=\{\mathbf{x}_1,\mathbf{x}_2,\ldots,\mathbf{x}_K\}\) denote the target fMRI sequence represented on the cortical surface (fsLR), with \(\mathbf{X}\in\mathbb{R}^{K\times N_v}\), where \(K\) is the number of fMRI frames and each frame \(\mathbf{x}_k\in\mathbb{R}^{N_v}\) captures cortical activation across \(N_v\) vertices. Let \(\mathbf{S}\in\mathbb{R}^{C\times T_s}\) denote the temporally aligned EEG recording, where \(C\) is the number of electrodes and \(T_s\) is the total number of EEG samples. Temporal alignment accounts for hemodynamic response delays following standard preprocessing protocols.

\noindent\textbf{Sequence-Level fMRI Organization.}
We denote a continuous fMRI window of length \(K_w\) as
\[
\mathbf{X}_{k_0:k_0+K_w-1} \in \mathbb{R}^{K_w \times N_v},
\]
where each \(\mathbf{x}_k\) is a vertex-level cortical map. Unlike prior frame-by-frame approaches, we jointly model the entire \(K_w\)-frame sequence to capture temporal continuity.

\noindent\textbf{Sequence-Level EEG Conditioning.}
We condition the generator on a sequence of EEG windows
\[
\mathbf{S}_{k_0:k_0+K_w-1} = \{\mathbf{s}_{k_0}, \ldots, \mathbf{s}_{k_0+K_w-1}\},
\]
where each fMRI frame \(\mathbf{x}_k\) is temporally aligned with a corresponding EEG window \(\mathbf{s}_k\in\mathbb{R}^{C\times W}\), and \(W\) is the window length in EEG samples.

Our goal is to learn a conditional generative prior
\(
p_\theta(\mathbf{X}\mid \mathbf{S}),
\)
that supports two tasks under a unified formulation:

\begin{enumerate}
\item \textbf{EEG-to-fMRI translation.} Given an aligned EEG segment \(\{\mathbf{s}_{k_0}, \ldots, \mathbf{s}_{k_0+K_w-1}\}\), reconstruct the corresponding vertex-level fMRI sub-sequence:
\[
\hat{\mathbf{X}}_{k_0:k_0+K_w-1} \sim p_\theta\big(\mathbf{X}_{k_0:k_0+K_w-1}\mid \mathbf{s}_{k_0:k_0+K_w-1}\big).
\]

\item \textbf{EEG-guided intermediate fMRI frame reconstruction (InterRecon).} For temporal consistency evaluation, we use sparse anchor frames \(\mathbf{y} = \{\mathbf{x}_{k-\Delta},\, \mathbf{x}_{k+\Delta}\}\) (\(\Delta\ge 1\)) along with their corresponding EEG windows \(\{\mathbf{s}_{k-\Delta}, \ldots, \mathbf{s}_{k+\Delta}\}\) to constrain the reconstruction of intermediate frames at arbitrary query indices \(\mathcal{K}^* = \{k^{(1)},\ldots,k^{(M)}\} \subset (k-\Delta, k+\Delta)\):
\[
\hat{\mathbf{X}}_{\mathcal{K}^*} \sim p_\theta\big(\mathbf{X}_{\mathcal{K}^*}\,\big|\,\{\mathbf{s}_{k-\Delta}, \ldots, \mathbf{s}_{k+\Delta}\},\;\mathbf{y}\big)
\]
\end{enumerate}

\subsection{Denoising Null-Space Diffusion  Transformer}

We formulate EEG-to-fMRI reconstruction as a conditional denoising diffusion transformer process that jointly models spatiotemporal fMRI sequences. Our framework builds upon the diffusion transformer architecture (Figure \ref{fig:framework}-a)), adapted to handle vertex-level cortical representations with sequence-level EEG conditioning, as shown in Figure \ref{fig:framework}.

\noindent\textbf{Spatiotemporal Tokenization and EEG Conditioning.}
To jointly model the \(K_w\)-frame fMRI sequence, we tokenize the spatiotemporal volume into \((K_w \times N_v)\) vertex-level tokens, where each vertex at each time frame constitutes an individual token with temporal positional encoding to distinguish frames. EEG features are extracted for the entire \(K_w\)-frame window using a temporal encoder \(f_{\phi}\):
\[
\mathbf{h}_{\text{EEG}} = f_{\phi}(\mathbf{S}_{k_0:k_0+K_w-1}),
\]
producing a sequence-level representation. During reverse diffusion, each vertex token is conditioned on \(\mathbf{h}_{\text{EEG}}\) through cross-attention mechanisms at every transformer layer, enabling the model to capture temporal dependencies guided by the EEG signal.

\noindent\textbf{Diffusion Forward Process.}
The forward diffusion process gradually perturbs fMRI frames \(\mathbf{x}^{(0)}\) over \(N\) timesteps by injecting Gaussian noise according to a variance schedule \(\{\beta_n\}_{n=1}^{N}\):
\begin{equation}
q(\mathbf{x}^{(n)} \mid \mathbf{x}^{(n-1)}) = \mathcal{N}(\mathbf{x}^{(n)}; \sqrt{1-\beta_n} \mathbf{x}^{(n-1)}, \beta_n \mathbf{I}),
\label{eq:forward_step}
\end{equation}
which equivalently samples \(\mathbf{x}^{(n)} = \sqrt{1-\beta_n}\mathbf{x}^{(n-1)} + \sqrt{\beta_n}\boldsymbol{\epsilon}\) with \(\boldsymbol{\epsilon}\sim \mathcal{N}(0,\mathbf{I})\). By the reparameterization property, we can directly sample noisy states from the clean data at any diffusion step \(n\):
\begin{equation}
q(\mathbf{x}^{(n)} \mid \mathbf{x}^{(0)}) = \mathcal{N}(\mathbf{x}^{(n)}; \sqrt{\bar{\alpha}_n} \mathbf{x}^{(0)}, (1-\bar{\alpha}_n) \mathbf{I}),
\label{eq:forward_direct}
\end{equation}
where \(\alpha_n = 1-\beta_n\) and \(\bar{\alpha}_n = \prod_{i=1}^{n}\alpha_i\).

\noindent\textbf{Conditional Diffusion Sampling and Training for Translation.}
The reverse process learns to denoise by training a neural network \(\boldsymbol{\epsilon}_\theta(\mathbf{x}^{(n)}, n, \mathbf{h}_{\text{EEG}})\) to predict the injected noise, conditioned on EEG features. The model is optimized via the denoising score matching objective:
\begin{equation}
\mathcal{L}_{\text{diff}} = \mathbb{E}_{n, \mathbf{x}, \boldsymbol{\epsilon}} \left[ \| \boldsymbol{\epsilon} - \boldsymbol{\epsilon}_\theta(\mathbf{x}^{(n)}, n, \mathbf{h}_{\text{EEG}}) \|^2 \right],
\label{eq:diffusion_loss}
\end{equation}
where \(\mathbf{x}^{(n)} = \sqrt{\bar{\alpha}_n}\mathbf{x}^{(0)} + \sqrt{1-\bar{\alpha}_n}\boldsymbol{\epsilon}\) with \(\boldsymbol{\epsilon}\sim\mathcal{N}(0,\mathbf{I})\). During inference, the reverse process recovers \(\mathbf{x}^{(n-1)}\) from \(\mathbf{x}^{(n)}\) using the predicted noise to estimate the posterior mean.

\noindent\textbf{Null-Space Sampling for InterRecon.}
To enable intermediate fMRI frame reconstruction from sparse anchor frames without additional training, we employ null-space constrained sampling, as shown in Figure \ref{fig:framework}-b). We model the sparse observation as a linear measurement \(\mathbf{y} = \mathbf{A}\mathbf{X}_{1:K_w}\), where \(\mathbf{A} = \mathrm{diag}(m_1, \ldots, m_{K_w})\) with \(m_k = 1\) for observed anchor frames and \(m_k = 0\) for missing frames. The solution is decomposed into range space (enforcing measurement consistency) and null-space (capturing residual variations):
\[
\mathbf{x}\equiv\underbrace{\mathbf{A}^{\dagger}\mathbf{A}\mathbf{x}}_{\text{range-space}}+\underbrace{(\mathbf{I}-\mathbf{A}^{\dagger}\mathbf{A})\mathbf{x}}_{\text{null-space}}.
\]
At each reverse diffusion step \(n\), we first obtain the clean-data estimate using the noise predictor:
\[
\mathbf{x}_{0\mid n} = \frac{1}{\sqrt{\bar{\alpha}_n}}\left(\mathbf{x}^{(n)} - \sqrt{1-\bar{\alpha}_n}\boldsymbol{\epsilon}_\theta(\mathbf{x}^{(n)}, n, \mathbf{h}_{\text{EEG}})\right),
\]
then apply range-null projection to enforce anchor consistency:
\[
\hat{\mathbf{x}}_{0\mid n}=\mathbf{A}^{\dagger}\mathbf{y}+(\mathbf{I}-\mathbf{A}^{\dagger}\mathbf{A})\,\mathbf{x}_{0\mid n},
\]
which guarantees \(\mathbf{A}\hat{\mathbf{x}}_{0\mid n}\equiv\mathbf{y}\). Finally, we update the diffusion state using the corrected estimate:
\[
\mathbf{x}^{(n-1)} = \frac{\sqrt{\bar{\alpha}_{n-1}}\beta_n}{1-\bar{\alpha}_n}\hat{\mathbf{x}}_{0\mid n} + \frac{\sqrt{\alpha_n}(1-\bar{\alpha}_{n-1})}{1-\bar{\alpha}_n}\mathbf{x}^{(n)} + \sigma_n\boldsymbol{\epsilon}',
\]
where \(\boldsymbol{\epsilon}'\sim\mathcal{N}(0,\mathbf{I})\). Through this EEG-guided null-space refinement, intermediate frames are reconstructed to be temporally consistent with anchor frames while preserving functional information encoded in the EEG signal. Importantly, this projection is applied only during sampling, requiring no retraining or architectural modification.

\subsection{Model Architecture}

\noindent\textbf{EEG Encoder.}
We employ a temporal convolutional encoder to extract discriminative features from EEG windows. Given an input EEG window \(\mathbf{s}_k \in \mathbb{R}^{C \times W}\), the encoder processes it through \(L\) stacked 1D convolutional blocks with stride-2 downsampling, progressively increasing feature channels while reducing temporal resolution. Each block applies convolution, batch normalization, and GELU activation. After global average pooling over the temporal dimension, a two-layer MLP projects to a \(d_e\)-dimensional embedding, yielding \(f_\phi(\mathbf{s}_k) \in \mathbb{R}^{d_e}\) that encodes the temporal patterns within the EEG window.

\noindent\textbf{Linear fMRI autoencoder.}
To enable efficient diffusion modeling while preserving the range-null decomposition property required by null-space sampling, we employ a \emph{linear} MLP-based autoencoder as an implementation trick. The encoder maps each fMRI frame \(\mathbf{x}_k \in \mathbb{R}^{N_v}\) to a lower-dimensional representation \(\mathbf{z}_k \in \mathbb{R}^{d}\) (\(d \ll N_v\)) via a linear projection, and the decoder symmetrically reconstructs the fMRI frame via the inverse linear mapping. Critically, the linearity ensures that the null-space projection \((\mathbf{I}-\mathbf{A}^{\dagger}\mathbf{A})\) remains well-defined and the range-null decomposition holds exactly in the compressed space. Both encoder and decoder are trained end-to-end with the diffusion transformer model to minimize reconstruction error, allowing the model to operate in a compact representation without sacrificing the mathematical structure required for null-space sampling.

\subsection{Training \& Inference}

\textbf{Training.} We train the model end-to-end on paired EEG-fMRI sequences using the denoising score matching objective (Eq.~\ref{eq:diffusion_loss}). The linear fMRI autoencoder is jointly optimized to minimize reconstruction error while preserving the range-null decomposition structure. Training details, including hyperparameters, are provided in the supplementary material.

\noindent \textbf{Inference for Reconstruction.} Given a test EEG segment \(\{\mathbf{s}_{k_0}, \ldots, \mathbf{s}_{k_0+K_w-1}\}\), we initialize \(\mathbf{x}^{(N)} \sim \mathcal{N}(0, \mathbf{I})\) and iteratively denoise using the learned reverse process with EEG conditioning to obtain the reconstructed fMRI sequence \(\hat{\mathbf{X}}_{k_0:k_0+K_w-1}\).

\noindent \textbf{Inference for InterRecon.} Given anchor frames \(\mathbf{y}=\{\mathbf{x}_{k-\Delta},\,\mathbf{x}_{k+\Delta}\}\) and corresponding EEG windows, we perform constrained sampling with null-space projection at each denoising step (as described above) to reconstruct intermediate frames at arbitrary query indices \(\mathcal{K}^*\subset (k-\Delta, k+\Delta)\), enabling EEG-guided frame reconstruction at any temporal position without retraining.

The complete inference procedure is summarized in Algorithm~\ref{alg:unified}, which unifies both reconstruction tasks through conditional null-space projection.

\begin{algorithm}[t]
\caption{Unified EEG-Conditioned fMRI Reconstruction}
\label{alg:unified}
\begin{algorithmic}[1]
\REQUIRE EEG sequence \(\mathbf{S}\), diffusion steps \(N\), [optional] anchor frames \(\mathbf{y}\), measurement operator \(\mathbf{A}\)
\ENSURE Reconstructed fMRI sequence \(\hat{\mathbf{X}}\)
\STATE Encode EEG sequence: \(\mathbf{h}_{\text{EEG}} \gets f_{\phi}(\mathbf{S})\)
\STATE Initialize noise: \(\mathbf{x}^{(N)} \sim \mathcal{N}(0, \mathbf{I})\)
\FOR{\(n = N\) \textbf{to} \(1\)}
    \STATE Estimate clean data: \\ \(\mathbf{x}_{0\mid n} \gets \frac{1}{\sqrt{\bar{\alpha}_n}}(\mathbf{x}^{(n)} - \sqrt{1-\bar{\alpha}_n}\boldsymbol{\epsilon}_\theta(\mathbf{x}^{(n)}, n, \mathbf{h}_{\text{EEG}}))\)
    \IF{InterRecon mode}
        \STATE \(\hat{\mathbf{x}}_{0\mid n} \gets \mathbf{A}^{\dagger}\mathbf{y} + (\mathbf{I}-\mathbf{A}^{\dagger}\mathbf{A})\mathbf{x}_{0\mid n}\)
    \ELSE
        \STATE \(\hat{\mathbf{x}}_{0\mid n} \gets \mathbf{x}_{0\mid n}\) \%Translation mode
    \ENDIF
    \STATE \(\mathbf{x}^{(n-1)} \gets \frac{\sqrt{\bar{\alpha}_{n-1}}\beta_n}{1-\bar{\alpha}_n}\hat{\mathbf{x}}_{0\mid n} + \frac{\sqrt{\alpha_n}(1-\bar{\alpha}_{n-1})}{1-\bar{\alpha}_n}\mathbf{x}^{(n)} + \sigma_n\boldsymbol{\epsilon}'\), \\
    where \(\boldsymbol{\epsilon}'\sim\mathcal{N}(0,\mathbf{I})\)
\ENDFOR
\RETURN \(\hat{\mathbf{X}} \gets \mathbf{x}^{(0)}\)
\end{algorithmic}
\end{algorithm}

\section{Experiments}
\label{sec:experiments}

\subsection{Setting}

\begin{table*}
\centering
\small

\caption{
\textbf{Comparison of fMRI Reconstruction Performance.} 
We compare our method against baseline approaches across different frame lengths and spatial regions (whole brain, visual cortex, and visual–audio cortex). 
Results are averaged over six subjects.
}
\label{tab:fmri_results}

\renewcommand{\arraystretch}{1.15}
\setlength{\tabcolsep}{9pt}

\resizebox{0.95\textwidth}{!}{
\begin{tabular}{c|c|ccc|ccc|ccc}
\hline
\textbf{Frame} & \textbf{Method} &
\multicolumn{3}{c|}{\textbf{Whole Brain}} &
\multicolumn{3}{c|}{\textbf{Visual}} &
\multicolumn{3}{c}{\textbf{Visual+Audio}} \\
& &
MSE $\downarrow$ & r $\uparrow$ & cos $\uparrow$ &
MSE $\downarrow$ & r $\uparrow$ & cos $\uparrow$ &
MSE $\downarrow$ & r $\uparrow$ & cos $\uparrow$ \\
\hline

\multirow{5}{*}{3}
& CNN-TC~\cite{liu2019convolutional}  & 0.302 & 0.808 & 0.832 & 0.212 & 0.823 & 0.873 & 0.217 & 0.836 & 0.875 \\
& CNN-TAG~\cite{calhas2022eeg} & 0.297 & 0.812 & 0.835 & 0.207 & 0.828 & 0.879 & 0.212 & 0.840 & 0.878 \\
& E2FNet~\cite{roos2025brainwaves}  & 0.288 & 0.818 & 0.833 & 0.204 & 0.831 & 0.880 & 0.205 & 0.844 & 0.882 \\
& E2FGAN~\cite{roos2025brainwaves}  & \textbf{0.280} & 0.819 & 0.840 & \textbf{0.196} & 0.832 & 0.882 & 0.203 & 0.846 & 0.884 \\
& Ours    & 0.282 & \textbf{0.822} & \textbf{0.847} & \textbf{0.196} & \textbf{0.834} & \textbf{0.886} & \textbf{0.200} & \textbf{0.848} & \textbf{0.887} \\
\hline

\multirow{5}{*}{10}
& CNN-TC~\cite{liu2019convolutional}  & 0.315 & 0.804 & 0.824 & 0.230 & 0.822 & 0.861 & 0.204 & 0.830 & 0.867 \\
& CNN-TAG~\cite{calhas2022eeg} & 0.309 & 0.810 & 0.829 & 0.225 & 0.826 & 0.865 & 0.199 & 0.835 & 0.871 \\
& E2FNet~\cite{roos2025brainwaves}  & 0.297 & 0.819 & 0.836 & 0.215 & 0.833 & 0.872 & 0.192 & 0.842 & 0.878 \\
& E2FGAN~\cite{roos2025brainwaves}  & 0.290 & 0.822 & 0.839 & 0.210 & 0.836 & 0.875 & \textbf{0.188} & 0.845 & 0.881 \\
& Ours    & \textbf{0.277} & \textbf{0.824} & \textbf{0.849} & \textbf{0.193} & \textbf{0.834} & \textbf{0.887} & 0.197 & \textbf{0.848} & \textbf{0.889} \\
\hline

\multirow{5}{*}{30}
& CNN-TC~\cite{liu2019convolutional}  & 0.322 & 0.800 & 0.820 & 0.232 & 0.818 & 0.858 & 0.208 & 0.828 & 0.867 \\
& CNN-TAG~\cite{calhas2022eeg} & 0.319 & 0.803 & 0.823 & 0.228 & 0.820 & 0.860 & 0.204 & 0.831 & 0.869 \\
& E2FNet~\cite{roos2025brainwaves}  & 0.316 & 0.806 & 0.826 & 0.225 & 0.823 & 0.863 & 0.201 & 0.833 & 0.872 \\
& E2FGAN~\cite{roos2025brainwaves}  & 0.314 & 0.808 & 0.828 & 0.222 & 0.825 & 0.865 & 0.199 & 0.835 & 0.874 \\
& Ours    & \textbf{0.281} & \textbf{0.819} & \textbf{0.845} & \textbf{0.195} & \textbf{0.829} & \textbf{0.884} & \textbf{0.200} & \textbf{0.844} & \textbf{0.885} \\
\hline
\end{tabular}
}
\end{table*}

\noindent \textbf{Dataset.} 
We evaluate our method on the CineBrain~\cite{gao2025cinebrain} dataset, which contains time-synchronized EEG-fMRI recordings from six subjects during naturalistic movie watching. For each subject, the data are split into non-overlapping training and testing sets with an 80\%/20\% split.  Importantly, for each subject, the testing set contains unseen video clips that are not present in the training set, ensuring within-subject evaluation. The training data for each subject has a duration of approximately 288 minutes, and the testing data covers about 72 minutes. 

\noindent \textbf{Data Preprocessing.}
EEG signals are recorded from 64 channels at 1000 Hz. Following standard practice, EEG signals are normalized channel-wise to the range (-1, 1). Due to the hemodynamic response delay relative to the underlying neural activity, we apply a temporal shift of 4s between each EEG window and its corresponding fMRI frame, consistent with the delay established in CineBrain~\cite{gao2025cinebrain}. The fMRI data have a temporal resolution of 0.8s (TR = 0.8s). Preprocessing follows the official CineBrain protocol, implemented using the standard fMRIPrep pipeline. The steps include motion correction, susceptibility distortion correction, slice-timing correction, co-registration to the structural T1-weighted image, and normalization to the HCP fsLR-32k grayordinate space. After preprocessing, each fMRI frame is represented by 91282 grayordinates, consisting of 32492 cortical vertices per hemisphere and 26298 subcortical vertices covering the thalamus, striatum, hippocampus, and other deep gray-matter structures.

\noindent \textbf{Metrics.}
Beyond measuring reconstruction fidelity with Mean Squared Error (MSE), we further employ neural-level quantitative metrics, including Pearson correlation coefficient (r) and Cosine Similarity. These metrics capture spatial correspondence between predicted and ground-truth fMRI responses, providing biologically meaningful insight into neural representation quality.

\noindent \textbf{Baselines.}
For dynamic fMRI frame reconstruction, we compare our method with the following representative volumetric baselines: (1) \textbf{CNN-TC}~\cite{liu2019convolutional}, a convolutional transcoder. (2) \textbf{CNN-TAG}~\cite{calhas2022eeg}, a graph-attention network. (3) \textbf{E2fNet}~\cite{roos2025brainwaves}, a U-Net based encoder-decoder framework. (4) \textbf{E2fGAN}~\cite{roos2025brainwaves}, an adversarial variant of E2fNet with a discriminator. For the intermediate fMRI frame reconstruction (InterRecon) setting, where the middle frames are reconstructed, we evaluate our method with classical widely-used \textbf{linear interpolation}.

\noindent \textbf{Implementation Details.}
The EEG input is first embedded into a 512-dimensional latent space, while each fMRI frame is projected to a 1024-dimensional representation. The diffusion backbone is implemented as a Transformer consisting of 6 layers with 8 attention heads and a hidden dimension of 1024. During training, we adopt 1000 timesteps with a linear noise schedule following standard diffusion settings. For inference, we employ DDIM~\cite{song2021denoising} for efficient sampling with 50 denoising steps. The model is trained for 200 epochs using the AdamW optimizer ($\beta_1=0.9$, $\beta_2=0.999$), with a learning rate of $1\times10^{-4}$ and a batch size of 32. For visual decoding, the reconstructed fMRI representations are further processed by CineSync-f~\cite{gao2025cinebrain} to generate corresponding videos. All experiments are conducted on a single NVIDIA A100 GPU.

\subsection{Result and Analysis}

In this section, we quantitatively examine the model’s spatiotemporal pattern recognition capability, evaluate its performance under various reconstruction scenarios.

\begin{table}[t]
\centering
\renewcommand{\arraystretch}{1.15}
\caption{
Comparison of interpolation and null-space strategies.
We compare \textit{linear interpolation}, \textit{our model without null-space sampling}, and our \textit{full model} that incorporates null-space sampling for enhanced temporal and structural consistency.
}
\label{tab:nullspace}
\setlength{\tabcolsep}{14pt}
\resizebox{0.49\textwidth}{!}{
\begin{tabular}{lccc}
\toprule
Method & MSE $\downarrow$ & $r$ $\uparrow$ & Cos $\uparrow$ \\
\midrule
Linear                 & 0.280 & 0.830 & 0.851 \\
Ours w/o null space    & 0.272 & 0.839 & 0.852 \\
Ours w/ null space     & \textbf{0.250} & \textbf{0.852} & \textbf{0.865} \\
\bottomrule
\end{tabular}
}

\end{table}

\begin{figure}
\centering
\includegraphics[width=0.48\textwidth]{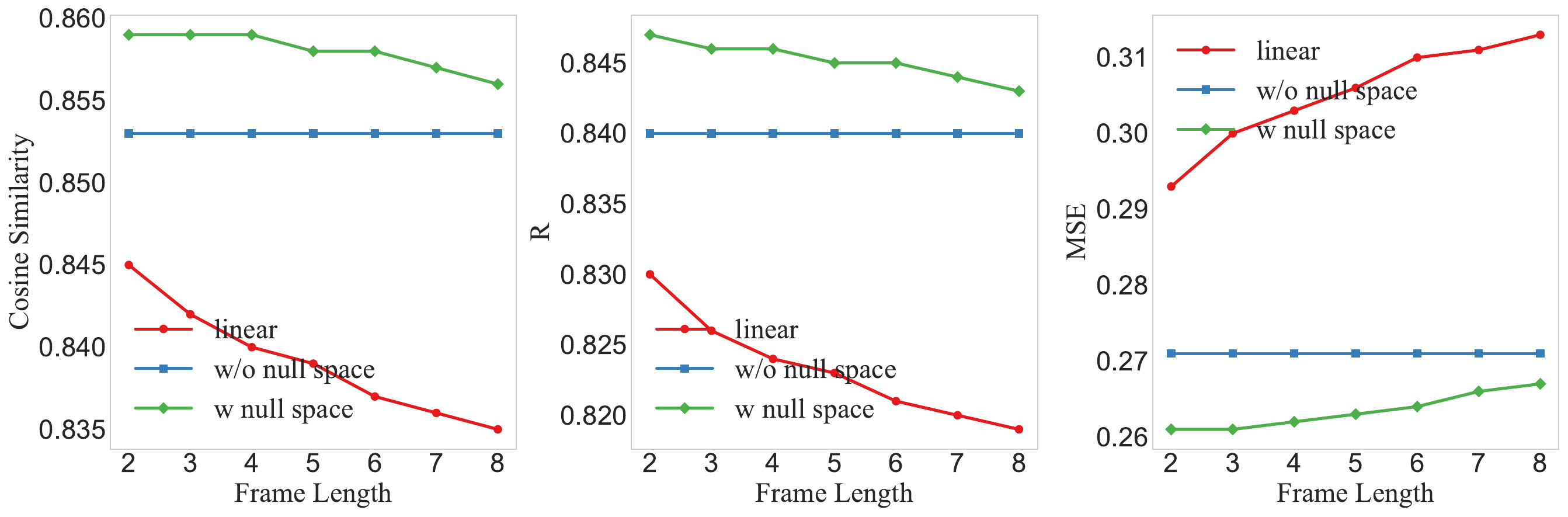}
\caption{Comparison of intermediate fMRI frame reconstruction (InterRecon) performance across different frame lengths and methods.}
\label{fig:frame_length}
\end{figure}

\begin{figure}
\centering
\includegraphics[width=0.99\columnwidth]{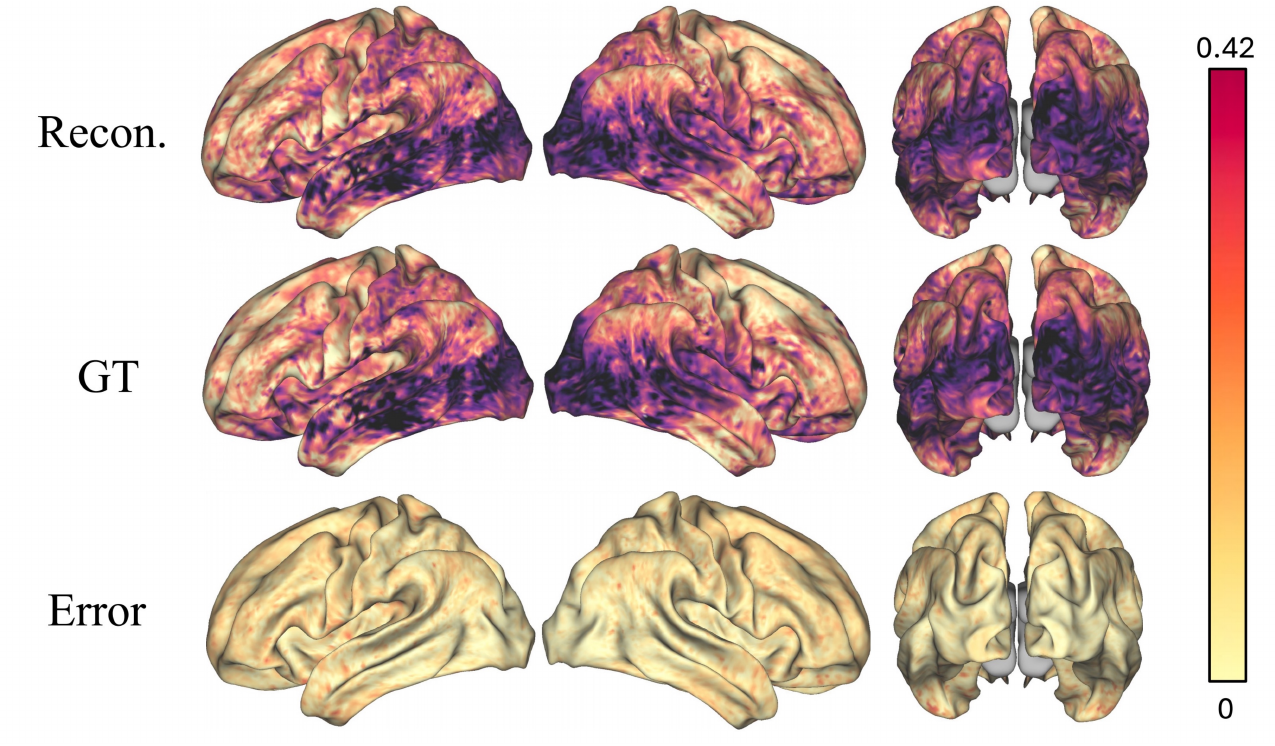}
\caption{Vertex-level cortical surface visualization of reconstructed fMRI alongside ground truth. The top row shows our reconstructed fMRI (Recon.); the middle row shows the ground-truth fMRI (GT); and the bottom row displays the reconstruction error map; the color scale indicates BOLD signal intensity.}
\label{fig:brain}
\end{figure}

\begin{figure*}
\centering
\includegraphics[width=0.97\textwidth]{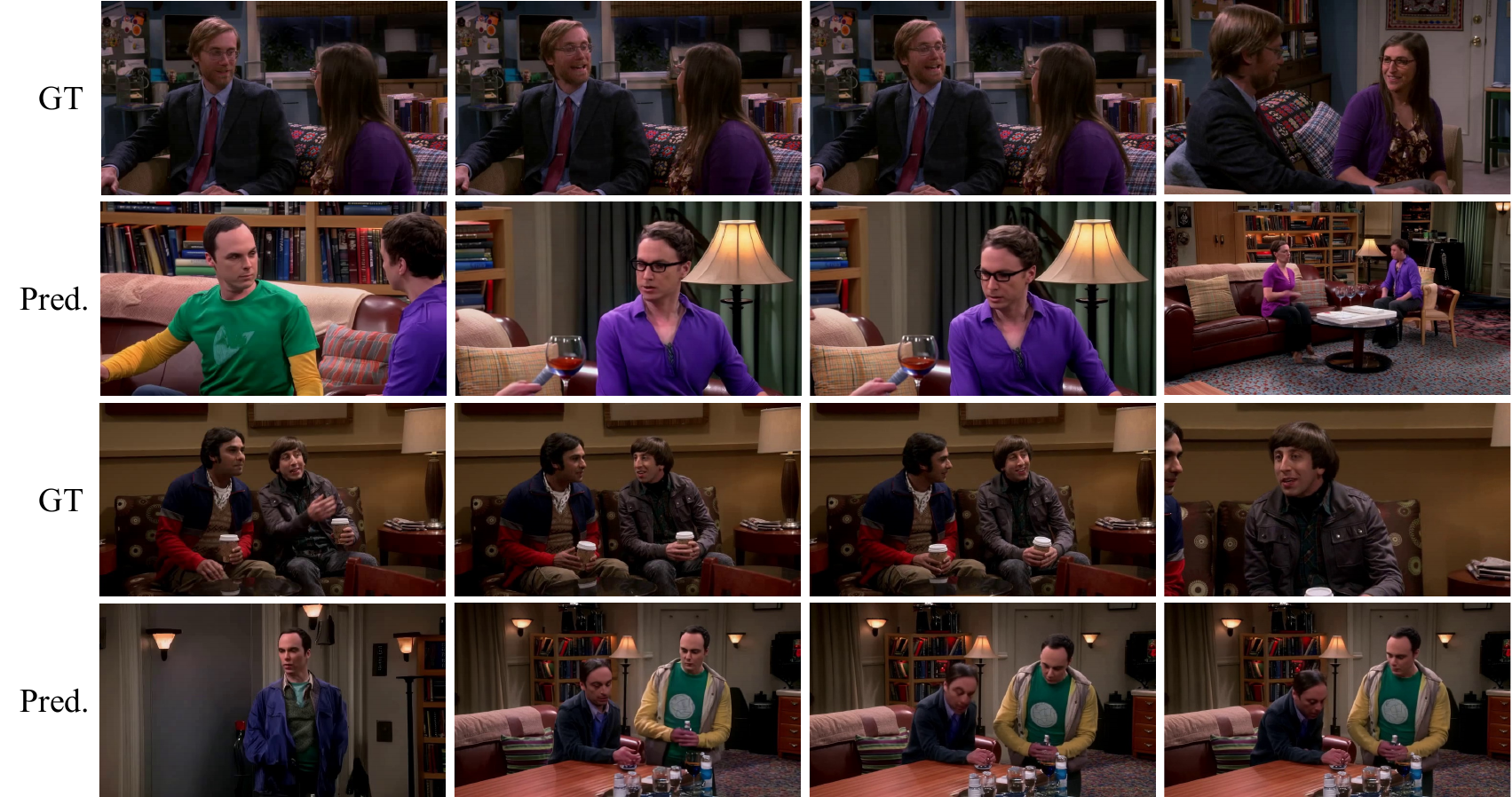}
\caption{Functional validation through visual decoding. Comparison between video frames decoded from our reconstructed fMRI (Pred.) and the corresponding ground-truth video frames (GT) using the CineSync-f decoder.}
\label{fig:movie}
\end{figure*}

\noindent\textbf{Dynamic fMRI Frames Reconstruction.} For spatial modeling, we conduct evaluations at three spatial regions of the fMRI representation. Whole brain: 91282 vertices; primary visual cortex (V1): 8405 vertices; combined visual auditory cortex (V1 + A1): 18946 vertices. For temporal modeling, we evaluate reconstruction under 3, 10, and 30 frame lengths to assess the model’s ability to recognize and reconstruct temporal patterns.

As shown in Table~\ref{tab:fmri_results}, across all spatial resolutions and frame lengths, our method consistently outperforms prior other methods, demonstrating strong spatiotemporal modeling capabilities.
\textbf{(1) Temporal robustness}. When increasing the length from 3 to 10 and 30 frames, competing methods such as CNN-TC, CNN-TAG, and E2F-based models show clear degradation in all metrics, whereas our model maintains highly stable performance across whole brain, V1, and V1+A1 regions. For example, in the whole brain region, our method preserves both low MSE (0.282 → 0.277 → 0.281) and high correlation (r: 0.822 → 0.824 → 0.819), indicating that temporal dependencies are effectively modeled even as temporal horizons increase. This stability highlights the model’s ability to recognize and propagate temporal patterns over short, mid, and long-range frame segments. \textbf{(2) Spatial robustness and whole-brain generalization.} Across the 91282 vertex whole-brain representation, our approach achieves the best cosine similarity for all frame lengths (0.847 / 0.849 / 0.845), demonstrating strong spatial consistency and global neural representation modeling. The ability to handle high-dimensional cortical mesh while maintaining superior performance indicates that the model captures stable and distributed cortical dynamics, rather than relying on localized patterns alone. \textbf{(3) Enhanced performance in task-relevant visual and audiovisual cortices.}
The performance in the visual (V1) and audiovisual (V1+A1) cortices is markedly higher than that at the whole-brain level. Notably, all metrics are consistently higher than the whole-brain results. Such specific region improvements are expected, as visual and audiovisual cortices are strongly driven by the movie-watching task. The superior performance in these biologically relevant areas indicates the model captures sensory-driven neural dynamics in a manner that aligns well with established neuroscience principles.

Overall, the results demonstrate that our framework delivers both temporal stability and spatial fidelity, scaling from short to long temporal windows while preserving accuracy across whole brain and task-relevant cortical areas. This provides strong evidence that our model not only reconstructs spatially plausible fMRI patterns but also effectively models the temporal evolution of neural activity.

\noindent \textbf{Intermediate fMRI Frames Reconstruction.} Table~\ref{tab:nullspace} shows the performance on subject 1 for generating the missing intermediate frame given its preceding and following fMRI frames. We compare three settings: a simple linear baseline, our model without null space sampling, and our full model with null space sampling. As shown, incorporating the null space constraint yields consistent improvements across all metrics. Figure~\ref{fig:frame_length} further examines the InterRecon setting across different frame lengths of subject 1. The linear interpolation baseline degrades noticeably as the temporal gap increases, whereas our null-space variant maintains much more stable performance and remains superior across all sequence lengths. Both the "w/o null space" and "w/ null space" results are produced without any additional training, using exactly the same checkpoint learned from the 10-frame dynamic reconstruction task. The comparison therefore, reflects solely the effect of enabling null-space sampling at inference. This demonstrates the flexibility of our diffusion-based framework, where the same model can be adapted to different reconstruction scenarios through sampling strategies alone. The superior performance of null space sampling indicates that explicitly constraining the solution space to respect boundary conditions leads to more biologically plausible intermediate frames that capture the smooth temporal evolution of neural activity.

\noindent \textbf{Visualization.} To assess the spatial accuracy of the reconstructed fMRI frames and evaluate their biological plausibility, we visualize the reconstructed cortical surface alongside the ground truth. The visualizations present lateral views of both hemispheres as well as a superior view, allowing inspection of large-scale spatial patterns across the cortex. As shown in Figure~\ref{fig:brain}, the reconstructed fMRI closely matches the ground truth in both intensity distribution and spatial topology, demonstrating high spatial fidelity with minimal reconstruction error across the entire cortical surface. These results indicate that our model captures coherent and anatomically consistent activation patterns, further supporting the biological plausibility of the generated fMRI signals.

\noindent \textbf{Downstream Visual Decoding Task.}
Beyond vertex-level reconstruction metrics, we validate the functional plausibility of reconstructed fMRI by evaluating whether it preserves task-relevant neural information for downstream visual decoding task. We employ CineSync-f~\cite{gao2025cinebrain}, a pretrained fMRI-to-video decoder, to reconstruct videos from our reconstructed fMRI frames. This provides a direct test of whether the reconstructed fMRI encodes biologically meaningful representations that support high-level cognitive tasks.

Qualitative results (Figure~\ref{fig:movie}) show that videos decoded from our reconstructed fMRI recover the coarse semantic structure of the original scenes. In the first case, the model correctly decodes a living-room setting with two characters engaged in conversation and captures salient elements such as the purple clothing. In the second case, the ground-truth frames (GT) depict a scene in which two male characters are seated on a couch and talking. The frames decoded from our reconstructed fMRI (Pred.) recover the overall scene layout, capturing the presence of the two male characters, their approximate poses, and the general conversational setting. Although finer details such as precise facial identity or exact clothing appearance are not perfectly reproduced, the reconstructions successfully preserve the overall scene context and high-level visual semantics. Moreover, the decoded frames form a temporally coherent sequence, reflecting the continuous nature of the underlying neural dynamics. These observations indicate that our reconstructed fMRI retains functionally meaningful information that supports downstream visual decoding, going beyond mere vertex-level similarity.

\section{Conclusion and Discussion}
\label{sec:conclusion.}
\textbf{Conclusion.} In this work, we formulate EEG-to-fMRI reconstruction simultaneously as a spatiotemporal pattern recognition problem, where the model extracts neurally meaningful dynamic features from EEG, and as a generative modeling task, where it reconstructs dynamic fMRI frames in high spatial resolution. Within a subject-specific setting, our diffusion-transformer-based framework reconstructs fMRI trajectories that are spatially detailed, temporally coherent, and consistent with known physiological dynamics. Our model achieves substantially better performance than previous representative baseline. To support intermediate frame reconstruction, we incorporate a null-space constrained sampling strategy that enforces consistency with observed frames during the generative process. This enables the model to infer missing temporal frames without retraining and provides a principled way to assess temporal consistency. Functional validation through task-driven visual decoding shows that the reconstructed fMRI frames preserve discriminative neural representations rather than only matching low-level similarity metrics.
Overall, this study demonstrates a feasible path to unify the temporal precision of EEG with the high spatial resolution of fMRI for dynamic brain activity modeling, cross-modal neural reconstruction, and future applications in neural decoding.

\noindent \textbf{Discussion.} While this framework achieves strong performance, two challenges remain. (1) fMRI responses during complex naturalistic tasks exhibit substantial cross-subject variability, making cross-subject modeling highly challenging. Our model is therefore trained within each subject, and extending it to subject-independent settings will require stronger anatomical or functional alignment. (2) We adopt a fixed EEG–fMRI delay following standard practice, yet the true hemodynamic latency is region-dependent and time-varying. Although our dynamic sequence modeling partially alleviates this mismatch, future work is needed to integrate adaptive or learnable alignment.

\newpage

\section*{Acknowledgments}
This project is supported by the National Key Research and Development Program of China (No. 2022YFC2405100).

{
    \small
    \bibliographystyle{ieeenat_fullname}
    \bibliography{main}
}

\newpage

\setcounter{section}{0}
\setcounter{figure}{0}
\setcounter{table}{0}
\renewcommand{\thesection}{S\arabic{section}}
\renewcommand{\thefigure}{S\arabic{figure}}
\renewcommand{\thetable}{S\arabic{table}}

\makeatletter
\let\arxiv@origsetcounter\setcounter
\renewcommand{\setcounter}[2]{%
  \def\arxiv@arg{#1}\def\arxiv@page{page}%
  \ifx\arxiv@arg\arxiv@page\else\arxiv@origsetcounter{#1}{#2}\fi
}
\makeatother

\setcounter{page}{1}

\begin{figure*}[!t]
    \centering
    \includegraphics[width=0.9\textwidth]{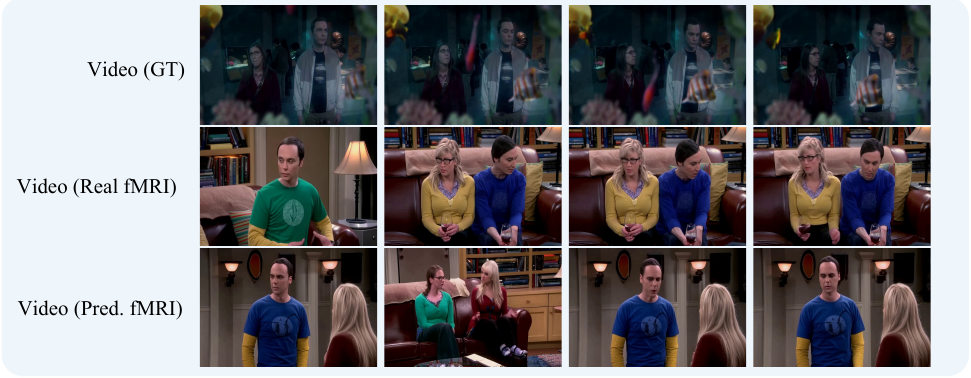}
    \caption{Functional validation through visual decoding. Comparison among the original stimulus video frames Video (GT), the frames decoded from ground-truth fMRI Video (Real fMRI), and the frames decoded from our reconstructed fMRI Video (Pred. fMRI) using the CineSync-f decoder. These three-way visual comparisons further illustrate the preservation of semantic content in the reconstructed fMRI.}
    \label{fig:movie}
    
    \end{figure*}

    \begin{figure*}[!t]
        \centering
        \includegraphics[width=0.9\textwidth]{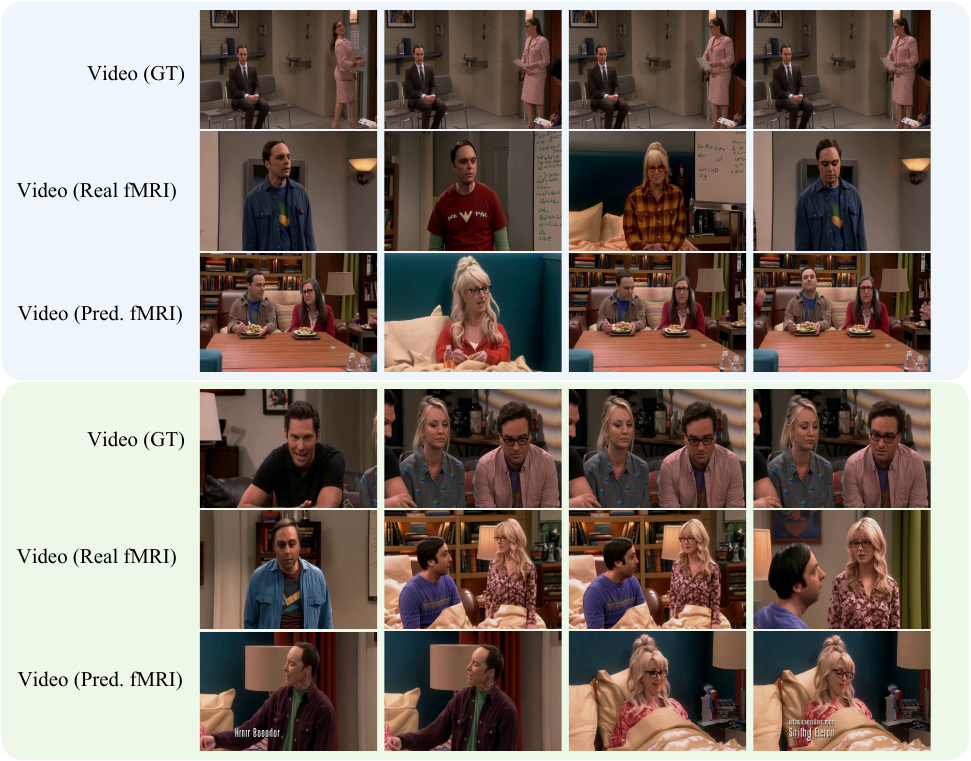}
        \caption{Functional validation through visual decoding. Comparison among the original stimulus video frames Video (GT), the frames decoded from ground-truth fMRI Video (Real fMRI), and the frames decoded from our reconstructed fMRI Video (Pred. fMRI) using the CineSync-f decoder. These three-way visual comparisons further illustrate the preservation of semantic content in the reconstructed fMRI.}
        \label{fig:movie2}
        
        \end{figure*}

\section{Data Preprocessing}
\label{sec:data_processing}

For CineBrain dataset, the fMRI preprocessing follows the CineBrain protocol and is performed using the standard fMRIPrep pipeline. The main steps include motion correction, susceptibility distortion correction, slice-timing correction, co-registration to the T1-weighted structural image, and normalization to the fsLR-32k grayordinate space defined by the Human Connectome Project (HCP). After preprocessing, each fMRI frame is represented by 91282 grayordinates, consisting of 32492 cortical vertices per hemisphere and 26298 subcortical gray-matter voxels covering thalamus, striatum, hippocampus, and other deep gray structures.

EEG data are acquired using an MRI-compatible 64-channel cap at a sampling rate of 1000~Hz, with simultaneous recording of ECG signals. Synchronization between EEG and fMRI is ensured by logging the exact fMRI TR timings during acquisition. EEG preprocessing follows the CineBrain protocol, with modifications to accommodate our recording configuration. A multi-step artifact removal procedure is applied to suppress scanner-induced and physiological noise while preserving neural activity. The preprocessing pipeline includes bandpass filtering between 0.1~Hz and 75~Hz to remove baseline drift and muscle artifacts, and a 50~Hz notch filter to attenuate powerline interference. ECG artifacts are first reduced using QRS-based correction methods, followed by independent component analysis (ICA) to isolate and remove residual artifacts. 
The recorded ECG signals are further used to refine artifact rejection through correlation-based adaptive adjustment, resulting in clean EEG data suitable for subsequent analysis.

\section{Subject-wise Dynamic fMRI Reconstruction}

To complement the averaged results reported in the main paper, we provide a full breakdown of dynamic fMRI reconstruction performance for all six subjects in the CineBrain dataset. Tables~\ref{tab:1}~\ref{tab:2}~\ref{tab:3} report the Mean Squared Error (MSE), Pearson correlation coefficient (r), and cosine similarity across three spatial regions: the whole brain (91282 vertices), the primary visual cortex V1 (8405 vertices), and the combined visual and auditory cortex V1+A1 (18946 vertices). Results are provided under three temporal window lengths of 3, 10, and 30 frames.

\begin{table}[t]
\centering
\caption{Subject-wise dynamic fMRI reconstruction performance (3-frame).}
\label{tab:1}
\renewcommand{\arraystretch}{1.25}
\setlength{\tabcolsep}{6pt}
\begin{tabular}{lccc}
\toprule
\textbf{Subject} & \textbf{MSE} & $\mathbf{r}$ & \textbf{Cos} \\
\midrule
sub1 & 0.272 $\pm$ 0.07 & 0.839 $\pm$ 0.04 & 0.852 $\pm$ 0.03 \\
sub2 & 0.189 $\pm$ 0.03 & 0.865 $\pm$ 0.03 & 0.890 $\pm$ 0.03 \\
sub3 & 0.285 $\pm$ 0.03 & 0.843 $\pm$ 0.06 & 0.850 $\pm$ 0.06 \\
sub4 & 0.346 $\pm$ 0.09 & 0.781 $\pm$ 0.08 & 0.815 $\pm$ 0.06 \\
sub5 & 0.288 $\pm$ 0.08 & 0.812 $\pm$ 0.04 & 0.839 $\pm$ 0.05 \\
sub6 & 0.314 $\pm$ 0.149 & 0.794 $\pm$ 0.06 & 0.834 $\pm$ 0.06 \\
\midrule
avg & 0.282 & 0.822 & 0.847 \\
\bottomrule
\end{tabular}
\end{table}

\begin{table}[t]
\centering
\caption{Subject-wise dynamic fMRI reconstruction performance (10-frame).}
\label{tab:2}
\renewcommand{\arraystretch}{1.25}
\setlength{\tabcolsep}{6pt}
\begin{tabular}{lccc}
\toprule
\textbf{Subject} & \textbf{MSE} & $\mathbf{r}$ & \textbf{Cos} \\
\midrule
sub1 & 0.271 $\pm$ 0.05 & 0.840 $\pm$ 0.03 & 0.853 $\pm$ 0.03 \\
sub2 & 0.188 $\pm$ 0.02 & 0.865 $\pm$ 0.03 & 0.890 $\pm$ 0.03 \\
sub3 & 0.284 $\pm$ 0.09 & 0.843 $\pm$ 0.04 & 0.850 $\pm$ 0.04 \\
sub4 & 0.342 $\pm$ 0.19 & 0.780 $\pm$ 0.07 & 0.816 $\pm$ 0.06 \\
sub5 & 0.275 $\pm$ 0.05 & 0.819 $\pm$ 0.03 & 0.846 $\pm$ 0.05 \\
sub6 & 0.303 $\pm$ 0.11 & 0.797 $\pm$ 0.04 & 0.838 $\pm$ 0.04 \\
\midrule
avg & 0.277 & 0.824 & 0.849 \\
\bottomrule
\end{tabular}
\end{table}

\begin{table}[t]
\centering
\caption{Subject-wise dynamic fMRI reconstruction performance (30-frame).}
\label{tab:3}
\renewcommand{\arraystretch}{1.25}
\setlength{\tabcolsep}{6pt}
\begin{tabular}{lccc}
\toprule
\textbf{Subject} & \textbf{MSE} & $\mathbf{r}$ & \textbf{Cos} \\
\midrule
sub1 & 0.283 $\pm$ 0.03 & 0.832 $\pm$ 0.02 & 0.846 $\pm$ 0.02 \\
sub2 & 0.196 $\pm$ 0.01 & 0.857 $\pm$ 0.02 & 0.883 $\pm$ 0.02 \\
sub3 & 0.290 $\pm$ 0.07 & 0.841 $\pm$ 0.03 & 0.849 $\pm$ 0.04 \\
sub4 & 0.344 $\pm$ 0.10 & 0.769 $\pm$ 0.04 & 0.807 $\pm$ 0.04 \\
sub5 & 0.280 $\pm$ 0.04 & 0.814 $\pm$ 0.02 & 0.842 $\pm$ 0.04 \\
sub6 & 0.294 $\pm$ 0.104 & 0.802 $\pm$ 0.03 & 0.841 $\pm$ 0.04 \\
\midrule
avg & 0.281 & 0.819 & 0.845 \\
\bottomrule
\end{tabular}
\end{table}

Furthermore, we report subject-wise performance for the InterRecon task. While the main paper provides results for Subject 1, here we include results for all six subjects. 


\section{Additional Video Reconstruction Results}

As shown in fig~\ref{fig:movie} and fig~\ref{fig:movie2}, we include additional visual comparisons of the original videos, the videos decoded from real fMRI signals, and those decoded from the reconstructed fMRI. These supplementary results provide further support for the quantitative and qualitative analyses discussed in the main manuscript.

Across diverse scenes and character configurations, the videos decoded from our predicted fMRI closely match those decoded from real fMRI. As shown in the figures, the decoding results based on reconstructed fMRI faithfully recover the overall scene layout, camera viewpoint, and background structure, while also capturing key aspects of the characters’ appearance, posture, interactions, and contextual semantics. Whether in indoor conversations, multi-person interactions, or rapidly changing scenes, the semantic content decoded from predicted fMRI is nearly equivalent to that obtained from real fMRI, indicating that our reconstructed fMRI successfully preserves the functional information required for neural visual decoding. In these varied scenarios, the EEG-driven fMRI reconstructions exhibit stable, high-level semantic representations that can be effectively read out by the visual decoder, yielding video outputs that are comparable to those produced using real fMRI. This functional consistency further demonstrates that the reconstructed fMRI carries meaningful representational value and can serve as a reliable substitute for real fMRI in visual cognition analyses.

\section{Additional Ablation on Latent Dimensionality}
\label{sec:latent_dim_ablation}

To further examine whether the linear autoencoder introduces a bottleneck effect, we perform additional ablation experiments with latent dimensionalities of 512, 1024, and 2048. As shown in Table~\ref{tab:latent_dim}, the downstream EEG-to-fMRI reconstruction performance remains stable across different latent dimensions, indicating that the proposed method is robust to this design choice and that the linear autoencoder is not the dominant limiting factor.

\begin{table}[t]
    \centering
    \caption{Ablation on latent dimensionality.}
    \label{tab:latent_dim}
    \renewcommand{\arraystretch}{1.1}
    \setlength{\tabcolsep}{10pt}
    \resizebox{\columnwidth}{!}{
    \begin{tabular}{lcccc}
    \toprule
    & \multicolumn{1}{c}{\textbf{LinearAE Reconstruction}} & \multicolumn{3}{c}{\textbf{EEGtofMRI}} \\
    \cmidrule(lr){2-2} \cmidrule(lr){3-5}
    \textbf{dim} & \textbf{MSE$\downarrow$} & \textbf{MSE$\downarrow$} & \textbf{r$\uparrow$} & \textbf{Cos$\uparrow$} \\
    \midrule
    512  & 0.0045 & 0.288 & 0.814 & 0.830 \\
    1024 & 0.0043 & \textbf{0.282} & \textbf{0.822} & \textbf{0.847} \\
    2048 & \textbf{0.0042} & 0.285 & 0.820 & 0.843 \\
    \bottomrule
    \end{tabular}}
    \end{table}

\makeatletter
\let\setcounter\arxiv@origsetcounter
\makeatother

\end{document}